\documentclass[aps,prb,twocolumn,showpacs,superscriptaddress]{revtex4}
\usepackage{graphicx,subfigure,amssymb,amsfonts,amsmath}

\begin{document}

\title{Quenched dynamics in interacting 
one-dimensional 
systems: Appearance of current carrying steady states from initial domain wall density profiles.
}
\author{Jarrett Lancaster}
\affiliation{Department of Physics, New York University, 4
Washington Place, New York, NY 10003 USA
}
\author{Emanuel Gull}
\affiliation{Department of Physics, Columbia University, 538 W.
120th Street, New York, NY 10027 USA}
\author{Aditi Mitra}
\affiliation{Department of Physics, New York University, 4
Washington Place, New York, NY 10003 USA
}
\date{\today}


\begin{abstract}
We investigate dynamics arising after an interaction quench in the quantum sine-Gordon model 
for a one-dimensional system initially prepared in a
spatially inhomogeneous domain wall state. We study the time-evolution of the density, 
current and equal time correlation functions using the
truncated Wigner approximation (TWA) to which quantum corrections are added in 
order to set the limits on its validity. For weak to moderate
strengths of the back-scattering interaction, the domain wall spreads out 
ballistically with the system within the light cone reaching a
nonequilibrium steady-state characterized by a net current flow.
A steady state current exists for a quench at the exactly solvable Luther-Emery point.
The magnitude of the current decreases with increasing strength of the back-scattering interaction. 
The two-point correlation function of the variable
canonically conjugate to the density reaches a spatially oscillating steady state 
at a wavelength inversely related to the current.

\end{abstract}


\maketitle

\section{Introduction}

A fundamental question in the study of strongly correlated systems concerns how a
quantum many-particle system prepared in an initial state which is not an exact eigenstate
of the Hamiltonian evolves in time, and under what conditions the system at long times
thermalizes as opposed to reaching a novel athermal state.~\cite{rev2010}
This question is particularly relevant now due to
experiments in cold-atomic gases which provide practical realizations of almost
ideal many-particle systems where the interaction between particles
and the external potentials acting on them can be changed rapidly in time.~\cite{rev2008}

Motivated by this, there has been considerable 
theoretical interest in studying the time-evolution of 
one-dimensional systems which are initially prepared in a spatially
inhomogeneous state by  
the application of external confining potentials. 
Nonequilibrium time-evolution is triggered when the external potentials are rapidly 
turned off which may be accompanied with a rapid change in the interaction between particles.  
For example,
the time-dependent density matrix renormalization group (TDMRG) has been used to study
the time-evolution of a domain wall in the XXZ spin-chain,~\cite{tdmrg1,tdmrg2} the conformal field
theory approach 
to study domain wall time evolution in the transverse-field Ising
chain at the gapless point,~\cite{CalabreseDW}
and the Algebraic Bethe Ansatz (ABA) 
to study Loschmidt echos for the XXZ chain
for an initial domain wall state.~\cite{Caux10a} 
ABA has also been used to study geometric quenches {i.e.}, the
time-evolution arising after two spatially separated regions have
been coupled together.~\cite{Caux10b}
The dynamics of hard-core bosons after an
initial confining potential was switched off was 
studied in Ref.~\onlinecite{Rigolqc}. Here  
it was found that the initial energy of
confinement resulted in the appearance of quasi-condensates at finite momentum.
Time-evolution of an initial density inhomogeneity after an interaction quench at the 
Luther-Emery point was studied in Ref.~\onlinecite{Foster10} where a power-law amplification
of the initial density profile was found. 

In this paper we study how a one-dimensional ($1$D) system prepared initially in a 
domain wall state corresponding to a density $\rho(x\rightarrow \pm \infty)=\pm \rho_0$,
(where $x$ is the coordinate along the chain, and $\rho_0$ is a constant)
evolves in time after a sudden interaction and potential quench.  
The $1$D system is modeled using the
quantum sine-Gordon (QSG) model which captures the low energy  physics of
a variety of one-dimensional systems such
as the spin-1/2 chain, interacting fermions with back-scattering interactions arising due to
Umpklapp processes, and
interacting bosons in an optical lattice.~\cite{Giamarchi}

The QSG model is integrable, its exact solution can be obtained using
Bethe-Ansatz.~\cite{BetheA} While this property has been exploited 
to a great extent to understand
equilibrium properties of many $1$D systems, extending Bethe-Ansatz
to study dynamics is a daunting task,
especially for the time-evolution of two-point correlation functions. 
Thus there is a necessity to develop approximate methods to study this model. 

Here we investigate the 
time-evolution of the QSG model semiclassically using the truncated Wigner approximation (TWA)
to which quantum corrections are added in order to set limits 
on its applicability.~\cite{Polkovreview}
Moreover our parameter regime corresponds 
to an interacting bose gas whose density is initially in the form of a domain wall.
We study how this initial state 
evolves in time as a result of a sudden switching on of an optical lattice, which
may or may not be accompanied by an interaction quench. An optical lattice is a source of
back-scattering interactions or Umpklapp processes, that tends to localize the bosons. 
Our aim is to understand how this physics affects the time-evolution of the domain wall
state. Note that domain walls like the one we study here 
have been created experimentally by subjecting equal mixtures of $^{87}Rb$ atoms in two
different hyperfine states to an external magnetic field gradient.~\cite{Weld09} Studying
quantum dynamics in such systems may soon be experimentally feasible.

One consequence of
quenched dynamics in integrable models is that the system often does not thermalize, 
with the long time behavior
depending non-trivially on the initial state. Here we find that an initial state in the form of
a domain wall evolves at long times into a current carrying state even in the presence of a 
back-scattering interaction of moderate strength. 
Moreover,
this net current flow has interesting consequences
for the behavior of two-point correlation functions. 
The lack of decay of current found here is consistent with the  
fact that the dc conductivity of a 1D system is infinite
even in the presence of back-scattering or Umpklapp processes.~\cite{Rosch00} 
The origin of the infinite conductivity is the large number of conserved quantities
in a 1D system, where some of them have a nonzero overlap with the 
current,~\cite{Rosch00,Zotos97} 
thus preventing an initial current carrying state from decaying to zero. 

We also justify the steady state current obtained from TWA by studying the
QSG model at the Luther-Emery point.  The Luther-Emery point 
is an exactly solvable point in the
gapped phase of the model. In particular we study how an initial current carrying
state evolves with time and find that a steady
state current (albeit of reduced magnitude) persists at long times. We also 
study how this current affects two-point correlation functions.

Since the QSG model is
a simplified model that neglects band-curvature and higher-order back-scattering or
Umpklapp
processes, an important question concerns 
to what extent it can capture quenched dynamics in realistic systems. 
The nonequilibrium time-evolution of the above domain wall initial state was
studied both for the exactly-solvable lattice model of the $XX$ spin chain, and 
its continuum counterpart, the Luttinger model.~\cite{Lancaster10a} 
The study of the density and various two-point correlation functions
revealed that both the lattice and the continuum model reached the same nonequilibrium
steady state, but differed in the
details of the time-evolution. Continuum theories
are far easier to handle both numerically and analytically than their lattice
counterparts. Therefore to what extent they can capture the steady state behavior
after a quench for general parameters is an   
open and important question which is beyond the scope of this paper. 
 
The paper is organized as follows. In section~\ref{TWA} we study the time evolution of
an initial domain wall state after a quench employing TWA. Results for
the density, current and two-point correlation functions are presented. In section~\ref{TWAqc}
we present results for the first quantum corrections to TWA for some representative 
cases and discuss the general applicability
of the TWA results.  In section~\ref{LE} we present
results for a quench at the exactly solvable Luther-Emery point for an initial
current carrying state. Here results for the
steady-state current as well as two-point correlation functions are presented.
Section~\ref{Conc} contains our conclusions.

\begin{figure}
\includegraphics[
width=0.95\columnwidth
]{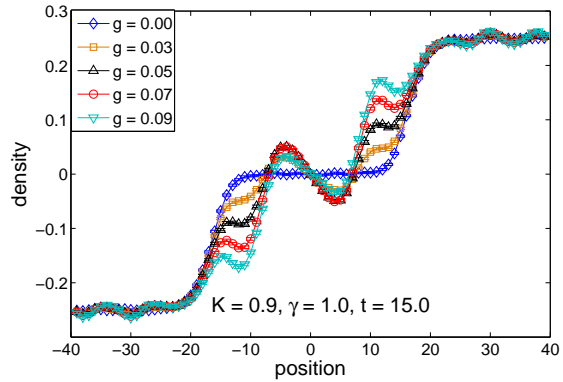}
\caption{(Color online) Density at time $t$=$15$ after the quench 
for $K$=$0.9$, $\gamma$=$1$ and several different $g$.
}
\label{szgcomp}
\end{figure}

\begin{figure}
\includegraphics[
width=0.95\columnwidth
]{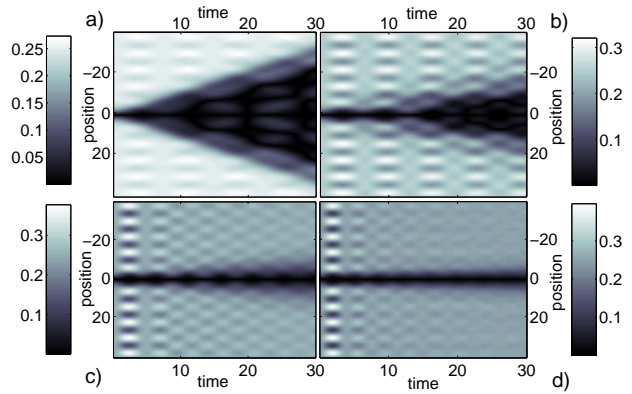}
\caption{(Color online) 
Contour plots for the magnitude of the density for $K$=$0.9$, $\gamma$=$1$ and for
values of $g$
a). g$=$0.05, b). g$=$0.2, c). g$=$0.6 and d). g$=$1.0. The density
at $t$=$0$ is $\rho(x)$=$(1/4)\tanh{(x/3)}$.}
\label{szcp}
\end{figure}

\begin{figure}
\includegraphics[
width=0.95\columnwidth
]{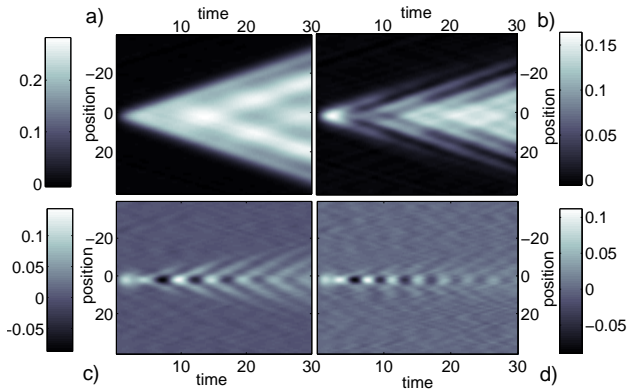}
\caption{(Color online) 
Contour plots for the -(current) for $K$=$0.9$, $\gamma$=$1$ and for
values of $g$
a). g$=$0.05, b). g$=$0.2, c). g$=$0.6 and d). g$=$1.0. The density
at $t$=$0$ is $\rho(x)$=$(1/4)\tanh{(x/3)}$
}
\label{icp}
\end{figure}

\begin{figure}
\includegraphics[
width=0.95\columnwidth]{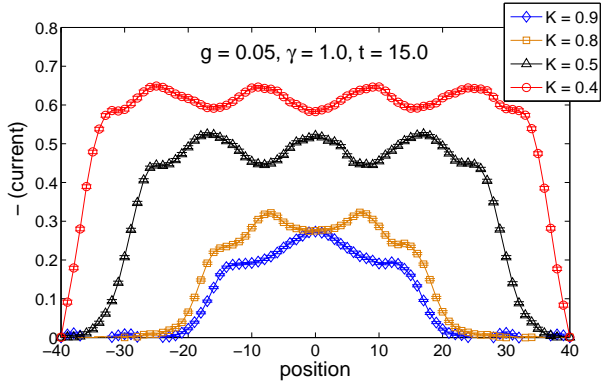}
\caption{(Color online) The current at $t$=$15$ for $g$=$0.05$, $\gamma$=$1$ and different $K$. 
}
\label{fig4a}
\end{figure}

\begin{figure}
\includegraphics[
width=0.95\columnwidth]{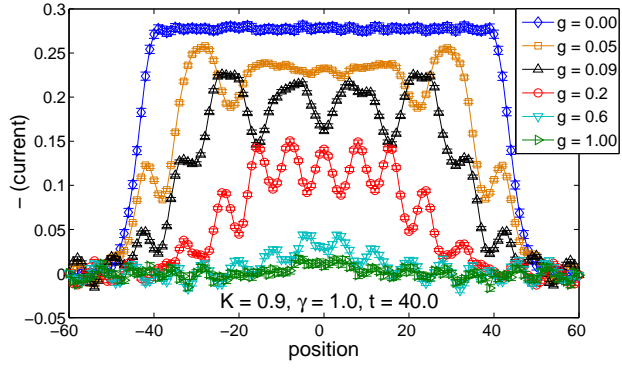}
\caption{(Color online) 
The current at $t$=$40\,$ for $K$=$0.9,\gamma$=$1\,$ and different $g$.  
}
\label{fig4b}
\end{figure}

\begin{figure}
\includegraphics[
width=0.95\columnwidth]{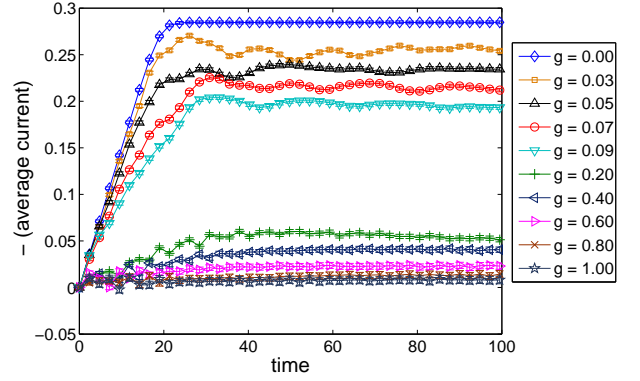}
\caption{(Color online) Time evolution of the current after spatially averaging over 
a strip of width $\delta x =40$ centered at $x=0$. }
\label{fig4c}
\end{figure}

\begin{figure}
\includegraphics[
width=0.95\columnwidth]{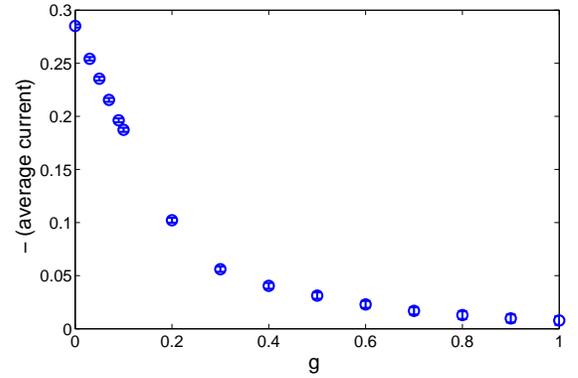}
\caption{(Color online) 
Dependence of the steady-state average current on interaction $g$ for $\gamma=1$ and
$K=0.9$. 
}
\label{fig4d}
\end{figure}

\section{Time-evolution using the Truncated Wigner Approximation} \label{TWA}

We start with an initial state which is the ground state of the Luttinger liquid, 
\begin{eqnarray}
H_i =&&
\frac{v_F}{2\pi}\int dx \left[\left(\partial_x \theta(x)\right)^2 
+ \left(\partial_x \phi(x)\right)^2\right. \nonumber \\
&&\left. +\frac{2}{v_F}h(x) \partial_x \phi(x)\right]
\label{Hi}
\end{eqnarray}
where in terms of bosonic creation and annihilation operators
$b_{p},b_{p}^{\dagger}$~\cite{Giamarchi},
\begin{eqnarray}
\phi(x) &=& -\frac{i\pi}{L}\sum_{p\neq0}\left(\frac{L|p|}{2\pi}\right)^{1/2}\frac{1}{p}
e^{-\alpha|p|/2-ipx}\left(b_{p}^{\dagger} + b_{-p}\right) \label{ft1}\\
\theta(x)
&=&\frac{i\pi}{L}\sum_{p\neq0}\left(\frac{L|p|}{2\pi}\right)^{1/2}
\frac{1}{|p|}e^{-\alpha|p|/2-ipx}\left(b_{p}^{\dagger} - b_{-p}\right)
\label{ft2}
\end{eqnarray}
and
$\left[\phi(x),\frac{1}{\pi}\partial_y\theta(y)\right]=i\delta(x-y)$. Above,
$v_F$ is the Fermi velocity or the velocity of the bosons, 
$\alpha$ a short-distance cut-off, $p$ the momentum, $L$ the length of the
system, and $h(x)$ is an external chemical-potential
which couples to the density $\rho(x)$=$-\frac{1}{\pi}
\partial_x \phi(x)$.
In the ground state of $H_i$ the density simply follows
the external field $\langle\rho(x)\rangle $=$ \frac{1}{\pi v_F}h(x)$.
We choose $h(x)$=$h_0\tanh{(x/\xi)}$ so that the initial density is a domain wall
of width $\xi$.
We study the case where at time $t$=$0$ the external field $h(x)$
is switched off. At the same time an optical-lattice is suddenly switched on which may  
be accompanied by an change in the interaction between bosons. Thus 
the time evolution for $t>0$ is due to the quantum sine-Gordon model,
\begin{eqnarray}
H_f=&&\frac{u}{2\pi}\int dx \left[K\left(\partial_x\theta(x)\right)^2
+ \frac{1}{K}\left(\partial_x \phi(x)\right)^2
\right] \nonumber \\
&&- g\int dx \cos\left(\gamma \phi(x)\right)\label{Hf}
\end{eqnarray}
Here $u$=$v_F/K$, $K$ being the
Luttinger parameter and $g$ the strength of the back-scattering interaction arising due to
a periodic potential. The ground state of $H_f$ has two well known 
phases,~\cite{Giamarchi} the localized 
(gapped) phase characterized by $\langle\phi\rangle\neq0$, 
and a delocalized (gapless) phase. 
The periodic potential is a relevant parameter for $2-\frac{\gamma^2 K}{4}>0$,
implying that the ground state has a gap for infinitesimally small $g$. 
On the other hand for $2-\frac{\gamma^2 K}{4}<0$,
a localized phase arises only for back-scattering strengths larger than a critical value
($g>g_c$). We will study quenched dynamics for parameters that are such
that $g$ is a relevant perturbation in equilibrium. Note that the initial domain
wall state is not an exact eigen-state of $H_f$. Neither is it 
related to the classical solitonic solution of the QSG model 
since the latter is a domain wall in the $\phi$ field,~\cite{Rajaraman} while our initial state
is a domain wall in $\partial_x\phi$.

When $g$=$0$, the time evolution of the system can be solved exactly.~\cite{Lancaster10a} 
For this case an initial density inhomogeneity shows 
typical light-cone dynamics~\cite{Calabrese} by spreading
out ballistically in either direction with
the velocity $u$, {\sl i.e.},
$\rho(x,t)$=$\frac{1}{2\pi v_F}\left[h(x+ut) + h(x-ut)\right]$. 
Since the system is closed, the energy is conserved. However 
during the course of the time-evolution, the energy density is 
transferred from the density to the current, the latter having the form 
\begin{eqnarray}
j(x,t)&=&\frac{1}{\pi}\frac{\partial \theta}{\partial x} \\
&=&\frac{1}{2\pi u K^2}\left[h(x-ut) - h(x+ut)\right]
\end{eqnarray}
In particular for 
$h(x)$=$h_0 \tanh(x/\xi)$, the energy density at 
$t$=$0$ is
${\cal E}$=$\frac{u\pi}{2K}\langle \rho(x)\rangle^2 \simeq h_0^2/(2\pi u K^3)$, while at long times, 
and for positions within the light-cone ($ut> |x|$ ) the energy density 
is ${\cal E}$=$\frac{u\pi K}{2}j^2$ where $j$=$-h_0/(\pi u K^2)$. 

Note that while any initial
density profile will give rise to transient currents, the special feature of a
domain wall density profile is that for a system of infinite length, the steady state
behavior is characterized by a net current flow. In particular for any finite time the 
current flows across a length $|x|=ut$ of the wire connecting the regions 
of high and low densities $\pm \rho_0= \pm h_0/(\pi v_F) $ at the two ends. 

We now explore how the time-evolution of the density, and the
long time behavior of the current and two-point correlation functions
is influenced by a back-scattering interaction ($g\neq 0$). 
The results are obtained using TWA which involves solving the classical
equations of motion with initial conditions weighted by the Wigner distribution function of
the initial state. Thus TWA is exact when $g$=$0$ while the effect of
$g$ is the leading correction in powers of $\hbar$.~\cite{Polkovreview}
Since $H_i$ is quadratic in the fields, it can be diagonalized by a simple shift, {\sl i.e.}, 
$H_i$=$\sum_{p\neq 0} v_F |p| a_p^{\dagger} a_p$, where $b_p$=$a_p + h_p/(v_F \sqrt{2\pi |p| L})
$, $h_p$ being the Fourier transform of $h(x)$.
The initial Wigner distribution function for the $a_p$ fields are Gaussian and are accessed by
a Monte-Carlo sampling. This is followed by a Fourier transform defined in Eqns.~\ref{ft1} and~\ref{ft2}
which gives the  
$\phi$ and $\theta$ fields at the initial time $t$=$0$. 
The classical equations of motion are then solved on a lattice up to a time $t$. All the data sets 
presented here are accompanied with error bars associated with the Monte-Carlo averaging. 
Lengths will be measured in
units of the lattice spacing $a$ which is also set equal to the short-distance cut-off $\alpha$.
Energy scales will be in units of $v_F/a$. The results will be presented for 
$h_0$=$\pi/4$ and an initial domain wall of width $\xi$=$3$.

\subsection{Time evolution of the density}

Fig.~\ref{szgcomp} shows the 
density at a time $t$=$15$ after the quench for $K$=$0.9$, $\gamma$=$1$ and several different $g$. 
The domain wall is found to
broaden with time with a velocity which is reduced from the velocity of
expansion $u$=$v_F/K$ when $g=0$. 
Moreover, unlike purely ballistic motion, the shape of the domain wall changes during the 
time-evolution.
The behavior of the density is clearer in the contour plots 
in Fig.~\ref{szcp}. For small $g$, the time-evolution shows a light-cone behavior
along with the appearance of spatial oscillations within the light-cone.
The amplitude of the oscillations
increase with $g$, while the wavelength of the oscillations is set by $\rho_0$.
Increasing $g$ gradually blurs the
light-cone, and eventually for very large $g$ the domain wall mass becomes so large that it hardly 
moves during the times calculated here. 

\subsection{Time evolution of the current}

The current behaves in a manner complementary to the
density and consistent with the continuity equation. 
Fig.~\ref{icp} shows contour plots for the current 
for parameters that are identical to that for
the density shown in Fig.~\ref{szcp}.
The current, like the density, 
fluctuates in space and time,
but on an average reaches a non-zero steady state within the light cone for
$g$ values that are not too large.
Fig.~\ref{fig4a}   
shows the current at time $t$=$15$ for a given $g$ and different $K$. As $K$ decreases,
the current increases as one expects from the analytic result for $g$=$0$. 
Fig.~\ref{fig4b} shows how the current behaves for  
a fixed $K$ and different $g$. Increasing $g$ not only reduces the overall 
velocity of expansion, but also reduces the magnitude of the current.
Fig.~\ref{fig4c} shows how the current spatially
averaged over a strip of width $40$ centered at the
origin evolves in time. There is a clear appearance of 
a current carrying steady state
whose magnitude decreases with $g$.
Note that the spatial averaging under-estimates
the time required to reach steady state as it under-estimates the amount
of current for $ut \leq  20$.

The dependence of the steady state current on $g$ is plotted in 
Fig.~\ref{fig4d} after time-averaging the current in Fig.~\ref{fig4c}
over a time window $t=40-100$ in order to eliminate the temporal fluctuations.
The steady state current is found to decrease linearly with $g$ for $g \ll 1$. 
Note that when $g\neq0$, the current does not commute with $H_f$. Yet the system
reaches a current carrying steady state. 
This is due to the fact that
the QSG model has a large number of other conserved quantities, some of which have a nonzero
overlap with the current operator, thus preventing the current to decay to zero. The lack
of decay of an initial current carrying state is also the origin of an infinite conductivity
in many integrable systems.~\cite{Giamarchi} 
It was argued in Ref.~\onlinecite{Rosch00} that
at least two different non-commuting Umpklapp processes are needed to violate conservation 
laws sufficiently so as to render the conductivity finite and thus cause the current 
to decay to zero.

\subsection{Steady-state correlation functions}

\begin{figure}
\includegraphics[
width=0.95\columnwidth
]{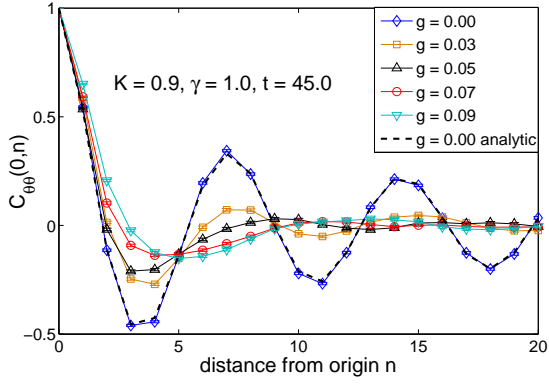}
\caption{(Color online) 
The equal time $C_{\theta\theta}(0t;nt)$ correlation function 
at $t$=$45$ for $K$=$0.9$, $\gamma$=$1$ and different $g$.
}
\label{cxxg}
\end{figure}

In this subsection we will study the following two equal time two-point correlation functions,
\begin{eqnarray} 
C_{\theta\theta}(xt;yt)&=&\langle e^{i\theta(x,t)} e^{-i\theta(y,t)}\rangle\\ 
C_{\phi\phi}(xt;yt)&=&\langle e^{i\phi(x,t)} e^{-i\phi(y,t)}\rangle
\end{eqnarray}
These are found to reach a 
nonequilibrium steady state for a time $ut >|x|,|y|$, {\sl i.e.}
for observation points that are within the light-cone.
The result for $C_{\theta\theta}$ 
for $K$=$0.9$ and different $g$ is plotted in Fig.~\ref{cxxg}. The TWA result for
$g$=$0$ is in agreement with the analytic 
result~\cite{Lancaster10a}
$C_{\theta\theta}(x,y;ut>|x|,|y|)$=$\exp\left[ih_0(y-x)/(v_F K)\right]
\left(\alpha/\mid x- y \mid\right)^{(1+K^{-2})/4}$. Thus when $g$=$0$, the correlation
function decays as a power-law 
with a slightly larger exponent than in equilibrium 
(the latter being 
$C^{eq}_{\theta\theta}(x,y)$=$\left(\alpha/\mid x- y \mid\right)^{1/(2K)}$). 
Moreover $C_{\theta\theta}$ shows
oscillations at wavelength $\lambda $=$\frac{2\pi v_F K}{h_0}
$=$ 2/j$, $j$ being the steady-state current within the light cone.
Note that the results for $h_0=0$ were obtained previously in Ref.~\onlinecite{Cazalilla06}
where the authors studied an interaction quench from a homogeneous initial state. 
The physical reason for the spatial oscillations when $h_0\neq 0$ is the
dephasing of the variable canonically conjugate to the density as the domain wall broadens.
This implies a dephasing of transverse spin components in the $XX$ spin chain
resulting in a spin-wave pattern at wavelength $\lambda$.~\cite{Lancaster10a} 
For a system of hard-core bosons, oscillations in $C_{\theta\theta}$
has the physical interpretation of the appearance of quasi-condensates at wave-vector 
$k$=$2\pi/\lambda$.~\cite{Rigolqc} 

\begin{figure}
\includegraphics[
width=0.95\columnwidth
]{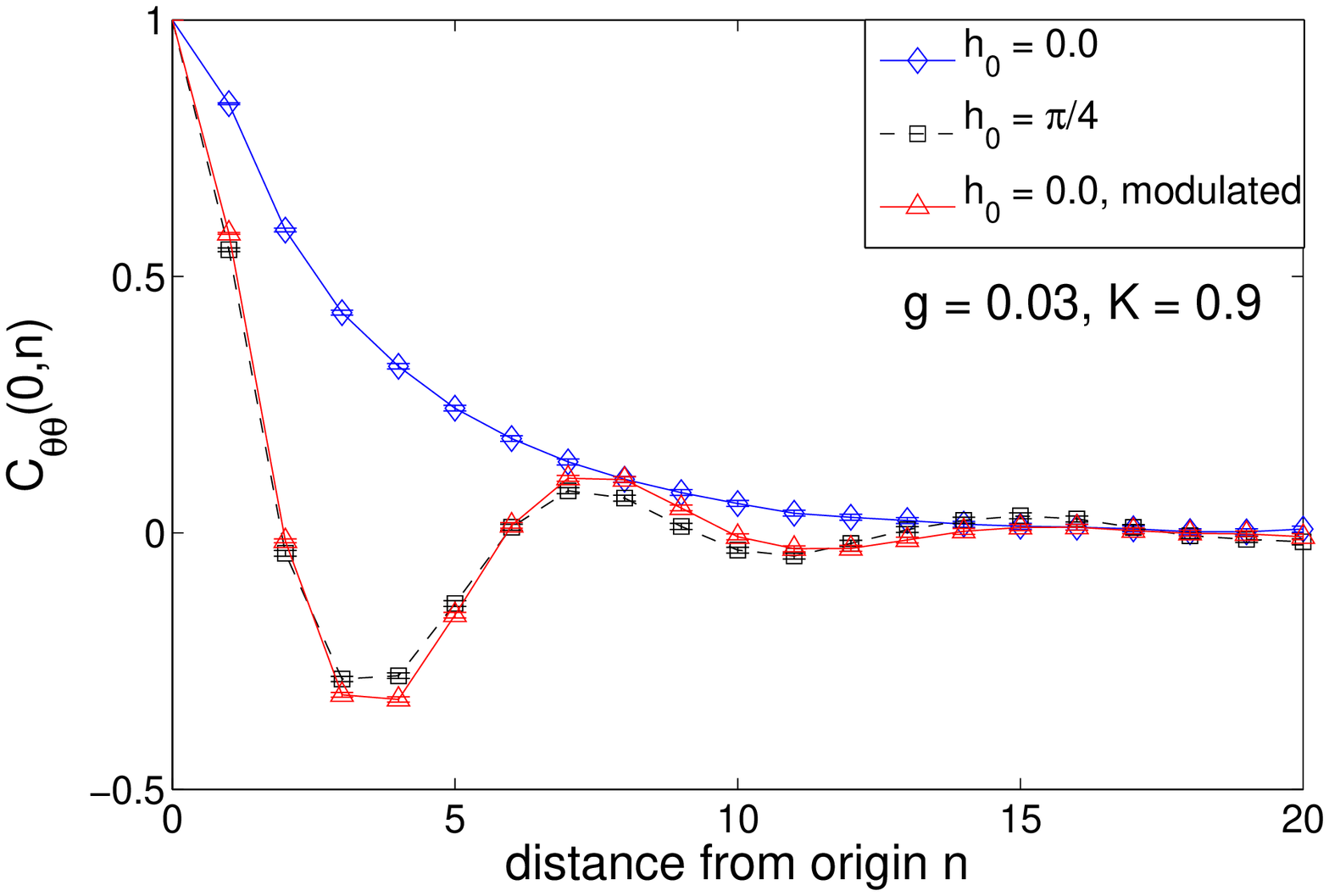}
\caption{(Color online) The equal time $C_{\theta\theta}(0t;nt)$ correlation function 
compared with correlation function for a homogeneous quench ($h_0$=$0$) and
modulated by $\cos(\pi j n)$ for $t$=$45$ and $g$=$0.03$, $K$=$0.9$, $\gamma$=$1$.}
\label{cxx0acomp}
\end{figure}

\begin{figure}
\includegraphics[
width=0.95\columnwidth
]{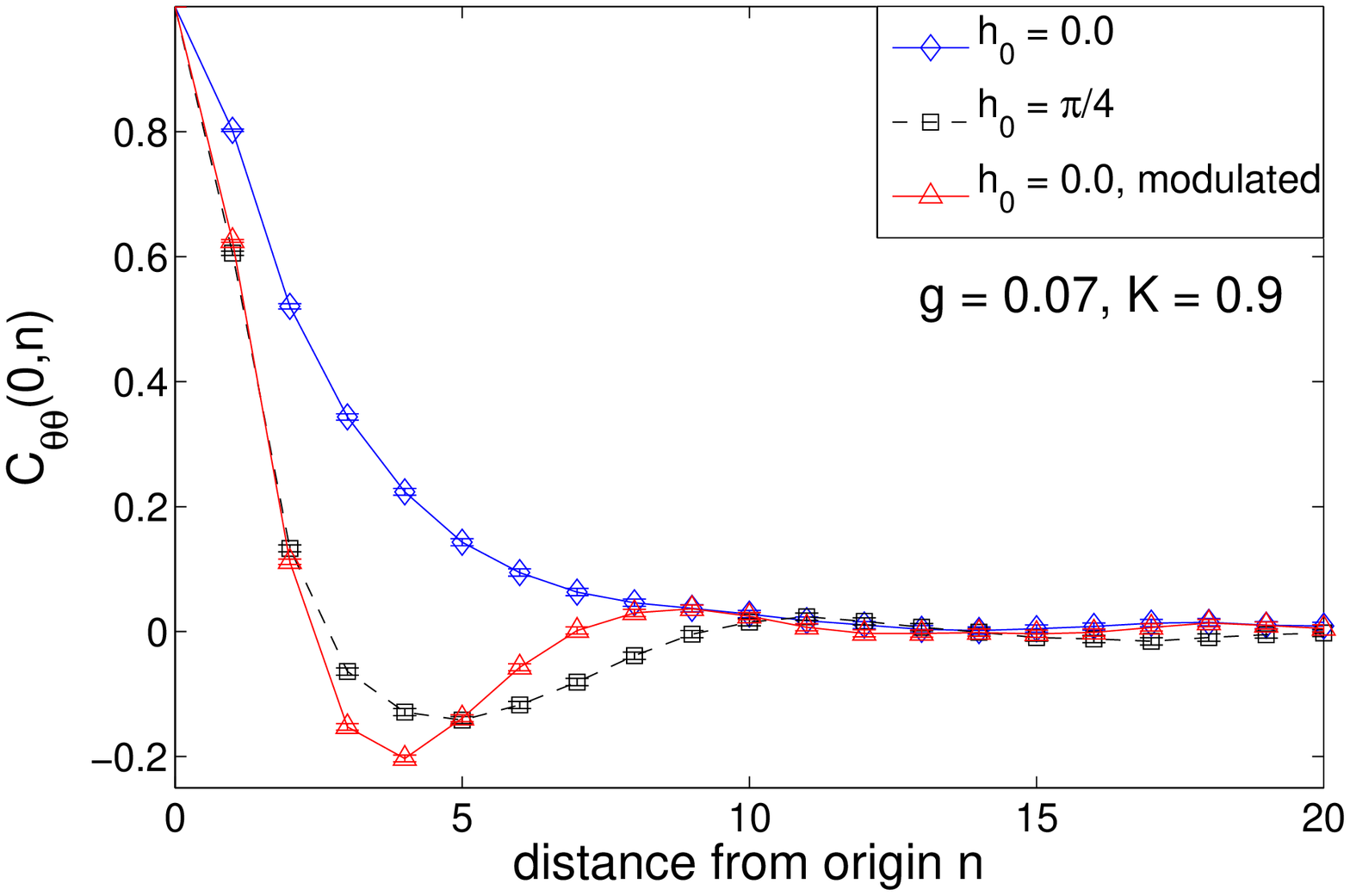}
\caption{(Color online) The equal time $C_{\theta\theta}(0t;nt)$ correlation function 
compared with correlation function for a homogeneous quench ($h_0$=$0$) and
modulated by $\cos(\pi j n)$ for $t=45$,$g$=$0.07$, $K$=$0.9$, $\gamma$=$1$.}
\label{cxx0bcomp}
\end{figure}

The TWA results presented here show that these effects can 
persist even in the presence of a back-scattering interaction, at least within
the continuum model.
In particular Fig.~\ref{cxxg} shows that when $g\neq 0$, $C_{\theta\theta}$ retains
the spatially oscillating form albeit at a wavelength that increases
with increasing $g$. Just as for $g$=$0$, one expects the current to set the 
wavelength of the oscillations. To check this
Figs.~\ref{cxx0acomp} and~\ref{cxx0bcomp} show a comparison between $C_{\theta\theta}(xt,yt)$ 
and $C_{\theta\theta}(h_0$=$0)(xt,yt)\cos(\pi j (x-y))$
where $C_{\theta\theta}(h_0$=$0)(xt,yt)$ is the correlation function
at long times after a homogeneous quench from
$H_i$ to $H_f$ ($h_0$=$0$ in $H_i$), while 
$j$ is the spatially averaged steady-state current in Fig.~\ref{fig4d}. The agreement
is found to be good at least for small $g$. 
The second effect of $g$ on $C_{\theta\theta}$
is to give rise to a faster decay in position. This decay also becomes faster 
for a given $g$ and on decreasing 
$K$ (not shown here) which takes the system deeper into 
the gapped phase of the QSG model.   

\begin{figure}
\includegraphics[
width=0.95\columnwidth
]{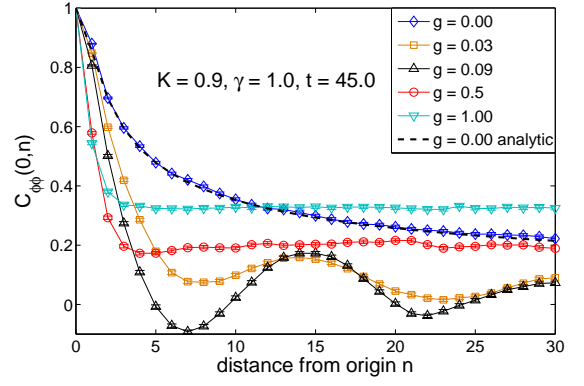}
\caption{(Color online) The equal time $C_{\phi\phi}(0t;nt)$ correlation function 
at $t$=$45$ and for $K$=$0.9$, $\gamma$=$1$ and different $g$. 
}
\label{czzg}
\end{figure}

Fig.~\ref{czzg} shows the behavior of $C_{\phi\phi}$ correlation function
after it has reached a steady state. The result for $g$=$0$ 
is~\cite{Cazalilla06,Lancaster10a}
$C_{\phi\phi}(x,y;ut >|x|,|y|)$=$\left(\alpha/\mid x- y\mid\right)^{(1+K^2)/4}$ which is also
characterized by a slightly faster power-law decay than in equilibrium (the latter being
$C^{eq}_{\phi\phi}(x,y)$=$\left(\alpha/\mid x- y\mid\right)^{K/2}$). Further unlike 
$C_{\theta\theta}$, $C_{\phi\phi}$ has
no memory of the initial spatial inhomogeneity for $g=0$. However this
is no longer the case for nonzero $g$.  
Fig.~\ref{czzg} shows that for small $g$, $C_{\phi\phi}$ can also show spatial oscillations.
Moreover, there is no appearance of
long-range order until $g$ is ${\cal O}(1)$, where the appearance of
the Ising gap corresponds to a nonzero asymptotic behavior of the two-point correlation
function. It is also consistent that the appearance of the gap in $C_{\phi\phi}$ coincides
with the value of $g$ for which the domain wall is almost static in Fig.~\ref{szcp}.

In equilibrium, the $\cos\gamma\phi$
interaction is a relevant perturbation for $2> \gamma^2 K/4$.~\cite{Giamarchi} 
Thus the $H_f$ parameters considered here are
those for which the ground state is the gapped Ising phase for $g$ of any strength. Yet there is no
signature of the gap in the quenched dynamics for $g \ll 1$. 
For this case the domain wall motion is ballistic,
and the $C_{\theta\theta}$ correlations persist over longer distances than in 
the gapped Ising phase. Similar observations have also been made in the study of
an interaction quench both in the bose-Hubbard model~\cite{Kollath08} and for 
a system of interacting  
fermions~\cite{manmana} where it was found that the system continued to 
show light-cone dynamics and gapless behavior
for parameters that correspond to the equilibrium gapped phase. 

The time-evolution of the density and current in the XXZ chain for an initial domain
wall state was studied in Ref.~\onlinecite{tdmrg1} 
employing TDMRG. There it was found that while a current
persists within the gapless phase, it decayed to zero in the gapped phase. This result
is different from what we find here where the current persists in the gapped 
phase as long as $g$ is not too large. There could
be two reasons for this difference. Firstly the parameters $\gamma$, $g$ and $K$ that we use here, do
not correspond to the parameters of the XXZ chain. Secondly, it is 
possible that the irrelevant operators that are not retained in 
the continuum model modify the long-time behavior, even though this was not found to 
be the case at the exactly solvable XX point ($J_z=0$, $K=1,g=0$).~\cite{Lancaster10a} 


\section{Quantum corrections to TWA} \label{TWAqc}

An important question concerns the validity of TWA. It was shown in Ref.~\onlinecite{Polkovreview} that 
in writing the time-evolution of an interacting system as a Keldysh path integral, TWA is the leading
correction in powers of $\hbar$. One may therefore check its validity by
expanding the path integral in higher powers of $\hbar$, and identify when these contributions 
become significant. We evaluate the first quantum correction along the lines of 
Ref.~\onlinecite{Polkovreview}.
Below we briefly outline the approach.

The expectation value of an observable $\hat{O}({\bf x}, {\bf p}, t)$ 
to leading order beyond TWA is~\cite{Polkovreview}
\begin{eqnarray}
&&\left\langle \hat{O}({\bf x},{\bf p},t)\right\rangle  \approx  
\int d{\bf x}_{0} d{\bf p}_{0} W_{0}({\bf x}_{0},{\bf p}_{0})\nonumber \\
&&\left[ 
1 - \int_{0}^{t}d\tau\frac{\hbar^{2}}{3!2!i^{2}}
\frac{\partial^{3}V}{\partial {\bf x}(\tau)^{3}}\frac{\partial^{3}}{\partial {\bf p}^{3}}
\right]
O_{W}({\bf x}, {\bf p},t)
\end{eqnarray}
where $W_{0}({\bf x}_{0}, {\bf p}_{0})\,$ is the initial Wigner distribution, 
and $O_{W}({\bf x}, {\bf p}, t) = \int d{\bf y} \left\langle {\bf x} - {\bf y}/2\right
| \hat{O}({\bf x},{\bf p},t)\left|{\bf x} + {\bf y}/2\right\rangle e^{i{\bf p}\cdot{\bf y}/\hbar}\,$ 
is the Weyl symbol of the operator $\hat{O}$. In the QSG model, ${\bf x}\rightarrow \phi(x)$, ${\bf p} 
\rightarrow \Pi(x) \equiv \frac{1}{\pi}\partial_{x}\theta(x)$, and 
$\int d{\bf x}d{\bf p} \rightarrow \int \mathcal{D}\phi(x)\mathcal{D}\Pi(x)$. 
In Ref.~\onlinecite{Polkovreview} the author implemented this correction by allowing 
a stochastic quantum jump in the momentum variable during the time evolution. 
This is done as follows: for each Monte Carlo step, we choose a set of initial conditions, 
weighted by the initial Wigner distribution, in accordance with TWA. For each 
set of initial conditions, we select a random position, $x_{n}$, and a random time, $\tau$. 
During the classical evolution, the field $\Pi_{n}(t) = \Pi(x_{n},t)$ is given a quantum 
kick at time $\tau\,$ by shifting $\Pi_{n}(\tau) \rightarrow \Pi_{n}(\tau) + \xi(\Delta\tau)^{1/3}$, 
where $\xi\,$ is a random weight chosen from a Gaussian distribution of zero mean and 
unit variance, and $\Delta\tau\,$ is a small time interval. Here, we take $\Delta \tau\,$ 
equal to the integration time step size, $\Delta t$. This process of 
sampling $\tau$, $x_{n}$, and $\xi$ is repeated for a given set of 
initial conditions. Thus the quantum correction to TWA is~\cite{Polkovreview}
\begin{equation}
\left\langle - \frac{tNg\gamma^{3}}{8}\sin[\gamma\phi_{n}(\tau)]
\left(\xi^{3}/3 - \xi\right)\hat{O}(\phi,\Pi,t)\right\rangle,
\end{equation}
where $N\,$ is the number of spatial points.

\begin{figure}
\includegraphics[
width=0.95\columnwidth
]{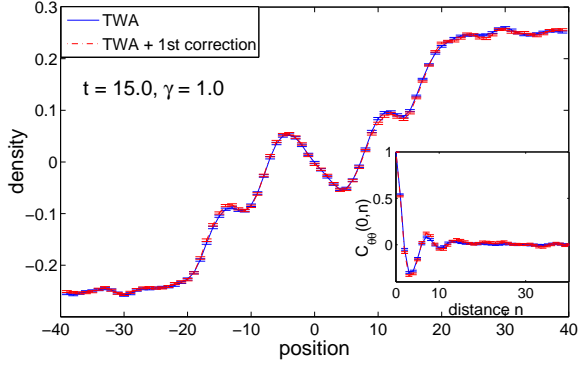}
\caption{(Color online) TWA and first quantum correction for 
the density (main panel) and equal time $C_{\theta\theta}$ correlation function (inset) 
for $K$=$0.9, g$=$0.05, \gamma$=$1$ and $t$=$15$. 
}
\label{szqc1a}
\end{figure}

\begin{figure}
\includegraphics[
width=0.95\columnwidth
]{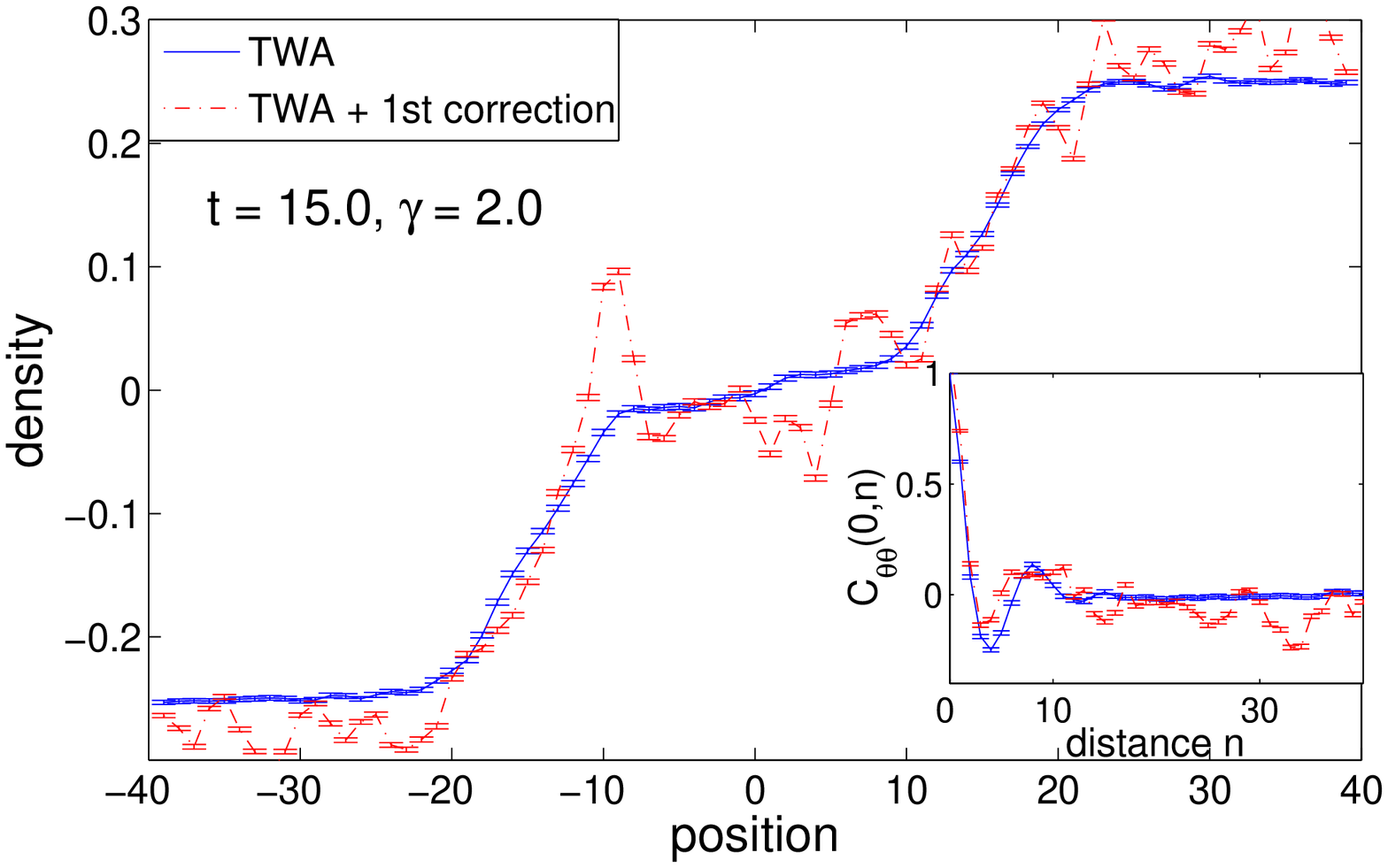}
\caption{(Color online)  TWA and first quantum correction for 
the density (main panel) and equal time $C_{\theta\theta}$ correlation function (inset) 
for $K$=$0.9, g$=$0.05, \gamma$=$2$ and $t$=$15$. 
}
\label{szqc1b}
\end{figure}

The results for the first quantum
correction for $\gamma$=$1$ and $\gamma$=$2$ are shown in Fig.~\ref{szqc1a} and Fig.~\ref{szqc1b}
respectively. 
As expected, the larger the coefficient $\gamma$, 
the larger the quantum fluctuations in the $\phi$ field,
causing TWA to break down sooner. We find TWA to work very well for 
$\gamma$=$1$ up to times $t=15$. On the other hand, for the same times, the
quantum corrections  for $\gamma$=$2$ are significant.

\begin{figure}
\includegraphics[
width=0.95\columnwidth
]{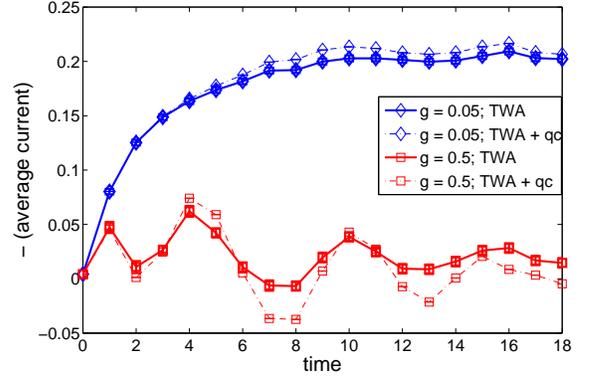}
\caption{(Color online)  TWA and first quantum correction to the current spatially averaged
over the light-cone $u t$ for 
$K$=$0.9$ and $\gamma$=$1$ and several different $g$. 
}
\label{iqc}
\end{figure}

It is also important to understand whether the steady-state current is a result of the
truncation scheme. To check this we plot the current evaluated from TWA along with the
first quantum correction in Fig.~\ref{iqc} for $\gamma=1$
and $K=0.9$. The current is now spatially averaged over the
non-interacting light-cone $|x| < u t$.  
The quantum correction is found to enhance the
current. This is expected on the grounds that TWA underestimates the quantum fluctuations, and 
therefore underestimates the extent of gapless behavior in the dynamics. 


\section{Quenched dynamics at the Luther-Emery point for an initial current carrying state}
\label{LE}

The main result of TWA was that a current carrying state can persist even in the
gapped phase of a model. 
In this section we will explore this physics at the Luther-Emery point of the
QSG model. In particular we will study 
how an initial current carrying state evolves in time
when the back-scattering interaction is suddenly switched on. We will also
explore the long time behavior of two point correlation functions.

The Luther-Emery point is a special point of the
QSG model where the  problem is rendered quadratic
after refermionization in
terms of left and right moving fermions $\psi_{L,R}$.~\cite{Giamarchi}
To see this, we rescale the fields in Eqn.~(\ref{Hf}) as $\phi' = \gamma\phi/2\,$ and 
$\theta' = 2\theta/\gamma$. Further if $K = 4/\gamma^2$, then $H_f$ may be written
as
\begin{eqnarray}
H_{f}^{\prime} & = & -iu\int dx \left[\psi_{R}^{\dagger}(x)\partial_{x}\psi_{R}(x) 
- \psi_{L}^{\dagger}(x)\partial_{x}\psi_{L}(x)\right]\nonumber\\
&+& 
m\int dx \left[ \psi_{R}^{\dagger}(x)\psi_{L}(x) + \psi_{L}^{\dagger}(x)\psi_{R}(x)\right]
\label{Hf'}
\end{eqnarray}
where $m=g\pi\alpha$ and 
\begin{eqnarray}
\psi_{R}(x) & = & \frac{\eta_{R}}{\sqrt{2\pi\alpha}}e^{-i[\phi'(x)-\theta'(x)]},\\
\psi_{L}(x) & = & \frac{\eta_{L}}{\sqrt{2\pi\alpha}}e^{i[\phi'(x)+\theta'(x)]},
\end{eqnarray}
$\eta_{R,L}\,$ are Klein factors to ensure the correct anticommutation 
relations among the fermions. 

We construct an  initial current carrying state which is
the ground state of the Hamiltonian 
\begin{eqnarray}
H_i^{\prime}=&&-iu\int dx \left[\psi_{R}^{\dagger}(x)
(\partial_x-i\frac{\mu}{u})\psi_{R}(x)\right. \nonumber \\
&&\left. - \psi_{L}^{\dagger}(x)(\partial_x -i\frac{\mu}{u})
\psi_{L}(x)\right]
\end{eqnarray}
where $2\mu$ is the chemical potential difference between right and left movers.  
We then study the time-evolution of this state for $t >0$ due to the Hamiltonian $H_f^{\prime}$
(Eq.~\ref{Hf'}). This Hamiltonian has a back-scattering interaction
of strength $m$, and no applied chemical potential difference between
right and left movers ($\mu=0$).

Defining $\psi_{R/L}(x)=\int \frac{dk}{2\pi}e^{ikx}\psi_{R/L}(k)$, 
the initial state is characterized by the occupations
\begin{eqnarray}
\langle \psi_R^{\dagger}(k)\psi_R(k)\rangle &=& \theta(-uk + \mu)\label{c1}\\
\langle \psi_L^{\dagger}(k)\psi_L(k)\rangle &=& \theta(uk -\mu)\label{c2}
\end{eqnarray}
The current is defined as 
\begin{eqnarray}
j(x)=u\left[\psi_R^{\dagger}(x)\psi_R(x)
-\psi_L^{\dagger}(x)\psi_L(x)
\right]
\end{eqnarray}
Thus the initial state is characterized by a current density
\begin{eqnarray}
j_0 = \mu/\pi
\end{eqnarray}
Since the theory is quadratic, the time-evolution can be studied in terms of
\begin{eqnarray}
\psi_R(k,t) &=& \psi_R(k)f(k,t) + \psi_L(k) g(k,t)\label{c3}\\
\psi_L(k,t) &=& \psi_L(k)f^*(k,t) + \psi_R(k) g(k,t) \label{c4}
\end{eqnarray}
where $f(k,t) = \cos(\omega_k t)- i\sin(\omega_k t)\cos(2\theta_k),
g(k,t) = -i \sin(\omega_k t)\sin{2\theta_k}$, $\omega_k$=$\sqrt{m^2 + u^2k^2},
\tan(2\theta_k)$=$m/(uk)$.

Using the above it is straightforward to work out the current at long times after the quench.
The current reaches a steady state  
\begin{eqnarray}
j=j_0 - (m/\pi)\tan^{-1}\left(j_0\pi/m\right)\label{jlt}
\end{eqnarray}
Note that this result is very similar to that obtained by TWA (Fig.~\ref{fig4d}) 
and predicts that the steady state current decays 
linearly with $m$ for $m\ll j_0$, while it decays as $1/m^2$
for large $m$.
The persistence of an initial  current carrying state
even in the gapped phase of a Hamiltonian was
also found in Ref.~\onlinecite{Klich}. Moreover in agreement with Ref.~\onlinecite{Klich}
we find that the
steady state current 
in the limit of very small initial
current $j_0\ll m$ is found to scale as the cubic power 
of the initial current $j \propto j_0^3$.
 
We now turn to the evaluation of the steady-state gap and two point correlation functions.
In terms of bosonic variables $\phi^{\prime} = \gamma \phi/2$ and $\theta^{\prime}= 2\theta/\gamma$, the
gap is
\begin{eqnarray}
\langle e^{2i\phi^{\prime}(x,t)}\rangle = -\langle \psi_R^{\dagger}(x)\psi_L(x)\rangle
\end{eqnarray}
while the basic two-point correlation functions are
\begin{eqnarray}
C_{\phi^{\prime}\phi^{\prime}}(x,t) 
&=& \langle e^{2i\phi^{\prime}(x,t)}e^{-2i\phi^{\prime}(0,t)}\rangle \nonumber \\
&=& \langle \psi_R^{\dagger}(xt)\psi_L(xt)\psi_L^{\dagger}(0t)\psi_R(0t)\rangle\\
C_{\theta^{\prime}\theta^{\prime}}(x,t) &=& 
\langle e^{-2i\theta^{\prime}(x,t)}e^{2i\theta^{\prime}(0,t)}\rangle \nonumber \\
&=& \langle \psi_R^{\dagger}(xt)\psi_L^{\dagger}(xt)\psi_L(0t)\psi_R(0t)\rangle
\end{eqnarray}
For long times after the quench we find
\begin{eqnarray}
\langle e^{2i\phi^{\prime}(x,t)}\rangle
= \frac{mu}{4\pi}\ln\left[\frac{u^2/\alpha^2}{m^2 +\pi^2j_0^2}\right]
= A(\alpha, m,j_0)
\label{gapLE}
\end{eqnarray}
where $\alpha$ is a short-distance cut-off. Thus the steady-state gap depends on the initial
current $j_0$.  

The two point correlations at long times are
\begin{eqnarray}
C_{\phi^{\prime}\phi^{\prime}}(x,t) &=& |A(\alpha,m,j_0)|^2+ |\frac{1}{2}\delta(x) + i I_b + I_a|^2\\
C_{\theta^{\prime}\theta^{\prime}}(x,t)
&=& \left(\frac{1}{2}\delta(x) + i I_b + I_a\right)
\left(\frac{1}{2}\delta(x) - i I_b - I_a\right) \nonumber \\
&-& \left(I_d+i I_c\right)^2
\end{eqnarray}
where 
\begin{eqnarray}
I_a &=& \int_0^{\mu/u}\frac{dk}{2\pi} \cos(k x)\frac{u^2k^2}{m^2 + u^2k^2}\\
I_b &=& \int_{\mu/u}^{\infty}\frac{dk}{2\pi}e^{-k\alpha}\sin{kx}\frac{u^2k^2}{m^2 + u^2k^2}\\
I_c &=& \int_0^{\mu/u}\frac{dk}{2\pi} \sin(k x)\frac{muk}{m^2 + u^2k^2}\\
I_d &=& \int_{\mu/u}^{\infty}e^{-k\alpha}\frac{dk}{2\pi}\cos{kx}\frac{muk}{m^2 + u^2k^2} 
\end{eqnarray}
For $j_0=0$ ($\mu =0 $), $C_{\phi^{\prime}\phi^{\prime}}$ reduces to the expression derived 
in.~\cite{Iucci} 
For $j_0\neq0$ and long distances $\mu x/u \gg 1, m x/u \gg 1$ we find,
\begin{eqnarray}
C_{\phi^{\prime}\phi^{\prime}}(x,t) 
= |A(\alpha,m,j_0)|^2 + \frac{1}{x^2}\left(\frac{\pi^2j_0^2}{\pi^2j_0^2 + m^2}
\right)^2
\end{eqnarray}
Thus the correlations are found to decay very slowly (as $1/x^2$) in position to
their long distance value of the square of the gap. This should be contrasted with
the equilibrium result where the decay to the long distance value is exponential.~\cite{Giamarchi} 
It is also interesting to compare this
result with that of an interaction quench from an initial state which
is the ground state of $H_i^{\prime}(\mu=0)$.~\cite{Iucci}
For this case the decay to the long distance value
is a power law $\left(1/x^6\right)$, but with a larger
exponent than found here for the current carrying state.  

The expression for $C_{\theta^{\prime}\theta^{\prime}}$ at long times after the quench is 
\begin{eqnarray}
C_{\theta^{\prime}\theta^{\prime}}(x,t\rightarrow\infty) = 
\frac{1}{x^2}\left(\frac{\pi^2 j_0^2}{m^2 + \pi^2j_0^2}\right)e^{-2i\pi j_0 x/u}
\end{eqnarray}
and shows a similar slow decay as $1/x^2$ in position (in contrast to an 
exponential decay in equilibrium). Moreover, the current flow imposes spatial oscillations
at a wavelength which is determined by the current.

This quench at the Luther Emery point did not involve a change in the Luttinger parameter $K$. 
Significantly different physics can occur after a similar quench that also changes 
the value of $K$. In Ref.~\onlinecite{Foster10}, 
the authors showed that changing $K\,$ can lead to the existence of ``super solitons'' 
at the Luther Emery point, where initial density inhomogeneities spread 
out with amplitudes that grow in time.

\section{Conclusions} \label{Conc}

In summary, we have performed a detailed study of quenched dynamics in an interacting $1$D system
prepared initially in a domain wall state. The model, being integrable, 
never thermalizes with the system 
reaching a nonequilibrium current carrying state which is robust even in the presence of 
moderate back-scattering interactions. 
The current has interesting consequences for the 
correlation functions, most notably the appearance of spatial oscillations
in the $C_{\theta\theta}$ correlation function.
Our predictions for the current 
can be tested 
experimentally using presently available one-dimensional optical lattice 
techniques.~\cite{rev2008,Weld09} 
\newline
{\it Acknowledgments:} AM is particularly indebted to A. Rosch and T. Giamarchi for 
helpful discussions. This work was supported by NSF-DMR (Award No. 0705584, 1004589 
for JL and AM, and 0705847 for EG).

\end{document}